\documentclass{elsart}
\usepackage{epsfig}
\usepackage{color}

\def\figbox#1;#2;{\parbox{#2cm}{\epsfig{file=#1.eps,width=#2cm}}}

\definecolor{dgreen}{rgb}{0,.5,0}

\definecolor{dmag}{rgb}{.6,0,.6}

\def\DAF{DA\char8NE} \def\epm{\ifm{e^+e^-}} 
\def\ifm#1{\relax\ifmmode#1\else$#1$\fi}

\def\ff{$\phi$--factory\ } \def\f{\ifm{\phi}}

\def\ab{\ifm{\sim}} \def\x{\ifm{\times}} 
\def\pt#1,#2,{\ifm{#1\x10^{#2}}} 
\def\brt{\hbox{BR(\ifm{\tau'})}}
\makeatletter 
\newdimen\z@ \z@=0pt 
\newskip\z@skip \z@skip=0pt plus0pt minus0pt 
\def\m@th{\mathsurround=\z@} 
\def\ialign{\everycr{}\tabskip\z@skip\halign} 
\def\eqalign#1{\null\,\vcenter{\openup\jot\m@th 
  \ialign{\strut\hfil$\displaystyle{##}$&$\displaystyle{{}##}$\hfil 
    \crcr#1\crcr}}\,} 
\makeatother 
\newcommand{\aff}[2]{Dipartimento di Fisica dell'Universit\`a #1 e Sezione INFN, 
#2, Italy.}
\newcommand{\affd}[1]{Dipartimento di Fisica dell'Universit\`a e Sezione INFN, 
#1, Italy.}

\newcommand{\eV}{{e\kern-.07em V}}
\newcommand{\MeV}{{\rm \,M\eV}}
\newcommand{\GeV}{{\rm G\eV}}
\newcommand{\ps}{{\rm \,ps}}

\newcommand{\mm}{{\rm \,mm}}

\newcommand{\m}{{\rm \,m}}
\newcommand{\um}{\ensuremath{\mathrm{\mu m}}}
\newcommand{\T}{{\rm \,T}}

\newcommand{\brn} {BR($K^\pm \rightarrow \pi^\pm \pi^0 \pi^0$)}

\begin{document}
\begin{frontmatter}
\title{Measurement of the branching ratio for the decay 
$K^\pm \rightarrow \pi^\pm \pi^0 \pi^0$ with the KLOE detector}
\collab{The KLOE Collaboration}
\author[Na]{A.~Aloisio},
\author[Na]{F.~Ambrosino},
\author[Frascati]{A.~Antonelli},
\author[Frascati]{M.~Antonelli},
\author[Roma3]{C.~Bacci},
\author[Roma3]{M.~Barva},
\author[Frascati]{G.~Bencivenni},
\author[Frascati]{S.~Bertolucci},
\author[Roma1]{C.~Bini},
\author[Frascati]{C.~Bloise}
% ...
\address[Bari]{\affd{Bari}}
\address[Beijing]{Permanent address: Institute of High Energy Physics of Academica 
Sinica, Beijing, China.}
\address[Frascati]{Laboratori Nazionali di Frascati dell'INFN, Frascati, Italy.}
\address[Karlsruhe]{Institut f\"ur Experimentelle Kernphysik, Universit\"at 
Karlsruhe, Germany.}
\address[Lecce]{\affd{Lecce}}
\address[Moscow]{Permanent address: Institute for Theoretical and Experimental 
Physics, Moscow, Russia.}
\address[Na]{Dipartimento di Scienze Fisiche dell'Universit\`a ``Federico II'' e 
Sezione INFN, Napoli, Italy.}
\address[Novosi]{Permanent address: Budker Institute of Nuclear Physics, 
Novosibirsk, Russia.}
\address[Pisa]{\affd{Pisa}}
\address[Energ]{Dipartimento di Energetica dell'Universit\`a ``La Sapienza'', 
Roma, Italy.}
\address[Roma1]{\aff{``La Sapienza''}{Roma}}
\address[Roma2]{\aff{``Tor Vergata''}{Roma}}
\address[Roma3]{\aff{``Roma Tre''}{Roma}}
\address[StonyBrook]{Physics Department, State University of New York at 
Stony Brook, USA.}
\address[Virginia]{Physics Department, University of Virginia, USA.}
\footnotetext[1]{Corresponding author: Andrea Ventura,
e-mail ventura@le.infn.it}
\footnotetext[2]{Corresponding author: Margherita Primavera,
e-mail primaver@le.infn.it}
\begin{abstract}

We have measured the absolute branching ratio \brn\ with the KLOE detector 
at the \DAF\ $e^+e^-$ collider. We collected \ab\pt3.3,7, tagged charged kaons, 
from the reaction $e^+e^-\rightarrow\phi\rightarrow K^+K^-$.
We find \brn=\pt(1.763\pm0.013_{\rm stat}\pm 0.022_{\rm syst}),-2,. 

\end{abstract}
\end{frontmatter}
Recently there has been renewed interest in three pion decays of charged 
kaons \cite{ambrosio}. Because of the small energy available in the reaction, 
$K\rightarrow 3\pi$ is an ideal process where to apply the notion of the 
Goldstone-boson nature of the pseudoscalar mesons, 
by testing the predictions obtained from the chiral lagrangian realization 
of the $\Delta S =1$ weak interactions \cite{maiani}.

The branching ratio for $K^\pm\rightarrow\pi^\pm\pi^0\pi^0$ 
($\tau'$$\ ^{3}$\footnotetext[3]{In the following text, this old notation will 
be often used to refer to $K^\pm\rightarrow\pi^\pm\pi^0\pi^0$ decay.}) decays 
is the least well known of the hadronic $K^\pm$ decay modes. The most accurate 
measurement to date of \brt, performed with a sample of \ab1300 
$K^+\to\pi^+\pi^0\pi^0$ decays in flight \cite{pdg,chiang}, has an accuracy of 
3.3\% and dates back to more than thirty years ago. 
We present a new determination of \brt, performed with the KLOE detector 
\cite{kloe} at the Frascati \ff\ \DAF\ \cite{Dafne}.
\DAF\ is an \epm\ collider operated at a CM energy $W=m_\f$\ab1020\ MeV/$c^2$.
About 50\% of the $\phi$ mesons decay to $K^+K^-$ pairs. Detection of a $K^\pm$ 
meson tags the presence of a $K^\mp$. Samples of $N_K$ (positive and negative) 
tagged kaons can be identified and used to search for the $\tau'$ decay. 
The branching ratio is then given by BR=$N(\tau')/N_K$. 
This procedure allows to determine directly the absolute branching ratio of 
interest. The one necessary condition is that the trigger of the {\it tagging} 
kaon should be only slightly dependent on the decay mode of the {\it tagged} kaon. 
We ensure this to be the case by verifying that the {\it tagging} kaon did in 
fact trigger by itself, using the complete set of information, recorded for 
each event, concerning all signals produced and processed by the trigger 
system~\cite{TRGnim}. 
The measurement described in the following is based on data collected at
the \f\ peak in 2001 and 2002, corresponding to an integrated luminosity 
of 441 pb$^{-1}$ or, equivalently, to the production of \ab\pt1.5,9, $\f$ mesons.

The KLOE detector consists of a large cylindrical drift chamber surrounded 
by a lead-scintillating fiber sampling calorimeter. A superconducting coil 
outside the calorimeter provides a 0.52\T\ magnetic field parallel to the beam 
axis. The drift chamber~\cite{DCnim} is 4\m\ in diameter and 3.3\m\ long. 
It uses aluminium field wires, a 90\% helium--10\% isobutane gas mixture and 
is constructed entirely with carbon fiber-epoxy composites. Transparency to 
photons is thus maximized and multiple Coulomb scattering minimized.
Single point resolutions are $\sim\!150\,\um$ in the transverse plane and 
$\sim\!2\mm$ longitudinally. Typical momentum resolution is 
$\sigma(p_T)/p_T\!\leq\!0.4\%$. Vertices are reconstructed with a spatial 
resolution of $\sim\!3\mm$. 

The calorimeter~\cite{EmCnim}, divided into a barrel and two endcaps, covers 
98\% of the solid angle. The readout granularity is \ab4.4$\times$4.4 cm$^2$ 
with 2440 ``cells'' arranged in five layers. Cells with signals close in time 
and space are grouped into a ``calorimeter cluster''. For each cluster, 
the energy deposit is the sum of the single cell energies. Arrival time 
and position are calculated as energy-weighted averages over the fired cells.
The energy resolution is $\sigma_E/E\!=\!5.7\%/\sqrt{E\ (\GeV)}$ and the time 
resolution $\sigma_t\!=\!54\ps/\sqrt{E\ (\GeV)}\oplus50\ps.$ 
The KLOE trigger~\cite{TRGnim} requires two local energy deposits in the 
calorimeter above threshold ($50 \MeV$ in the barrel, 150 MeV in the endcaps) 
or an appropriate hit multiplicity in the chamber. The trigger system also
produces a cosmic-ray veto using the signals from the outer calorimeter layers.

Charged kaons are identified by observation of their two-body decays: 
$\mu^\pm\nu_\mu\ (K^\pm_{\mu2})$ and $\pi^\pm\pi^0\ (K^\pm_{\pi2})$, comprising 
\ab85\% of all $K^\pm$ decays. The monochromatic charged particle momentum in 
the kaon CM produces a very clean signature and is exploited to identify $K^\pm$
production and thus tag the presence of another charged kaon, whose decay can be 
investigated. 

\begin{figure}
\vbox to6cm{
\parbox{5.cm}{\caption{Decay particle momentum in the kaon CM, in the $\pi^\pm$ 
mass hypothesis, for data and Monte Carlo. The distributions are normalized to 
unity. \label{daumom}}}
\parbox{5cm}{\includegraphics{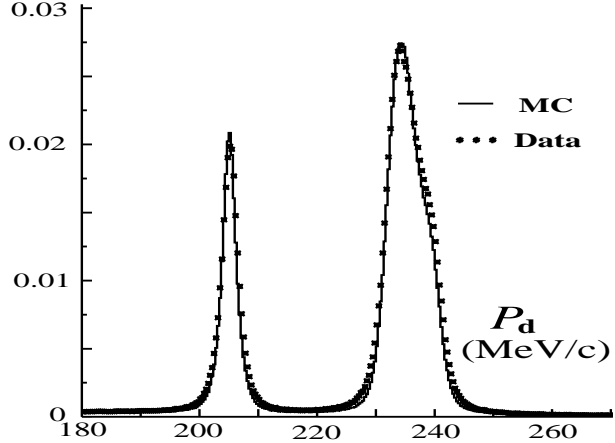}}}
\end{figure}

As mentioned above, since all information on how the trigger was formed is 
available, we require that the tagging kaon indeed satisfied the trigger 
requirements. Charged kaon decays are found by looking for a charged particle 
originating from the interaction point and ending in a decay vertex reconstructed 
inside the drift chamber. 
We require that: 1) the radial distance of the decay vertex from the beam axis be 
between 40 and 150 cm; 2) the kaon momentum at the decay point be between 70 
and 130 MeV/$c$ and the point of closest approach, $x_c,y_c,z_c$, of its track 
to the beam axis satisfy $\sqrt{x^2_c+y^2_c}<10$ cm and $|z_c|<20$ cm (the origin 
of the reference frame is the collider interaction point); 3) the momentum of the 
decay product $p_d$, assumed to be a pion, satisfy $180<p_d<270$ MeV/$c$ in the 
kaon CM. In Fig. \ref{daumom} the $p_d$ distribution is shown for data and Monte 
Carlo: the two well distinct peaks correspond to $K^\pm_{\pi2}$ and $K^\pm_{\mu2}$ 
decays. 

$K^\pm_{\pi2}$ decay identification requires a decay particle momentum in the $K$ 
meson rest frame compatible with the expected value of 205.14 MeV/$c$ and detection 
of a $\pi^0$. The latter requires observation of two energy clusters in the 
calorimeter, photons, with energies $>$15 MeV and times of flight consistent with 
the decay vertex, which we refer to as ``on-time'' clusters in the following. 
In addition, we require $85< m_{\gamma\gamma}<185$ MeV/$c^2$. 
$K^\pm_{\mu2}$ decay identification requires the decay particle momentum in the 
$K$ meson rest frame to be compatible with 235.53 MeV/$c$. Furthermore, the missing 
mass $m_m$ at the decay vertex must satisfy $|m^2_m|<5000$ MeV$^2$/$c^4$, which 
removes residual background from $\pi^\pm\pi^0$ decays. 
The main background in the two samples is mostly due to $K^\pm_{e3}$ and 
$K^\pm_{\mu 3}$ decays, especially for $K^\pm_{\pi2}$ decays.
From Monte Carlo simulation we find: 
$f_{\rm bckgnd}(\pi^\pm\pi^0) = (0.37\pm 0.02)\% ,\ f_{\rm bckgnd}(\mu^\pm\nu)
 = (0.21\pm 0.02)\%$,
where the quoted errors include statistical and systematic contributions.

$\tau'$ decays are events in which a decay vertex is observed for the {\it tagged} 
kaon, as for the {\it tagging} kaon above. In addition, at least four {\it on-time} 
energy deposits with $E\!>\!15$ MeV must be detected. The maximum charged pion 
momentum in the $K^\pm$ frame is $132.95$ MeV/$c$ for $\tau'$ decays. We accept 
only pions with $p_{\pi^\pm}\!<\!135$ MeV/$c$. 
The on-time condition ensures that the arrival times are consistent with photons 
originating at the same point, the decay vertex. Since the cluster time $t_i$ is 
well known but the decay instant is not, we construct the differences $\Delta_{jk}$ 
between the quantities $t_i-r_i/c$ for all cluster pairs $j,k$, with 
$j,k=1,\dots ,N$, $k>j$ and $r_i$ the distance from the centroid of the cluster $i$ 
to the decay vertex. Each difference is normalized using the appropriate error so 
that its variance is unity:
$$\Delta_{jk}={t_j-t_k-(r_j-r_k)/c\over\sqrt{\sigma_{t,j}^2+\sigma_{t,k}^2}} \ .$$
We require $|\Delta_{jk}|<4$ for all pairs.
Accidental on-time background clusters are effectively rejected by this cut. 
Events with more than four on-time clusters in the final sample, apart from the 
decay $K^\pm\rightarrow\pi^\pm\pi^0\pi^0\gamma$ (BR$<10^{-5}$), are due to residual 
on-time background and photon showers reconstructed as multiple clusters. 
We finally require that $\sum E_i<450$ MeV, for the four most energetic energy 
deposits. Data from 188 pb$^{-1}$ are used to search for $\tau'$ decays. The 
remainder of the data is divided into three subsets for the purposes of efficiency 
evaluation. We find 41896 $K^+$ and 41155 $K^-$.
We get 52253 and 30798 $\tau'$ decays which pass all requirements above, tagged by 
\pt1.9925,7, $K^\pm_{\mu2}$ and \pt1.2753,7, $K^\pm_{\pi2}$ decays, respectively.

As background in the $K^\pm\rightarrow\pi^\pm\pi^0\pi^0$ sample, we have 
considered $K^\pm _{\pi2}$, $K^\pm_{e3}$ and radiative decays like 
$K^\pm\to\pi^\pm\pi^0\gamma$, in which a ``spurious'' cluster in the calorimeter 
has been paired with the cluster of the radiated $\gamma$ (the probability for this 
to happen has been estimated from a $K^\pm_{\pi2}$ data sample to be $\sim 8\%$). 
The nuclear interaction of a charged kaon (mainly $K^-$) with the beam pipe and the
chamber walls can produce secondaries that simulate the $\tau'$ signal. 
Finally, we include $K^\pm\rightarrow l^\pm\pi^0\pi^0\nu_l$ ($K_{l4}'$) decays. 
The relative contributions of these backgrounds have been estimated by Monte Carlo 
simulation, while the total amount of background events has been determined by 
fitting the observed missing mass ($m_m$) spectrum at the decay vertex (assumed to 
be $K\to\pi+x$) with the sum of the simulated signal and background spectra. The 
background fraction estimated from the Monte Carlo simulation had to be corrected 
by a factor $1.13\pm 0.07$. After applying this correction we find that the average 
total background fraction is $f_{\rm bckgnd}(\tau')=(0.75\pm 0.11)\%$, where the 
quoted error includes the contributions from the finite statistics of the Monte 
Carlo spectra and from the fit. Contamination from other decays not originating 
from $K^\pm$ is negligible.

The overall $\tau'$ efficiency $\epsilon(\tau')$ is given by the product 
$\epsilon=\epsilon_K\epsilon_v\epsilon_\gamma$. $\epsilon_K$ is the efficiency to 
reconstruct the track of the charged kaon that undergoes $\tau'$ decay. 
$\epsilon_v$ includes the efficiency for decay vertex finding and losses due to 
the pion momentum cut. 
$\epsilon_\gamma$ is the efficiency for observing at least four photons in the 
calorimeter satisfying the energy and timing requirements above. Each factor has 
been measured separately for positive and negative kaon charge and for 
$K^\pm_{\mu2}$ and $K^\pm_{\pi2}$ tags by using special control samples 
\cite{pipinote} extracted from data taken in run periods close in time but 
different from the one used for signal selection. Control samples selected using 
calorimeter variables have been used to compute the efficiencies involving the 
drift chamber and vice-versa. 

\begin{figure}
\vspace{6.4cm}
\includegraphics{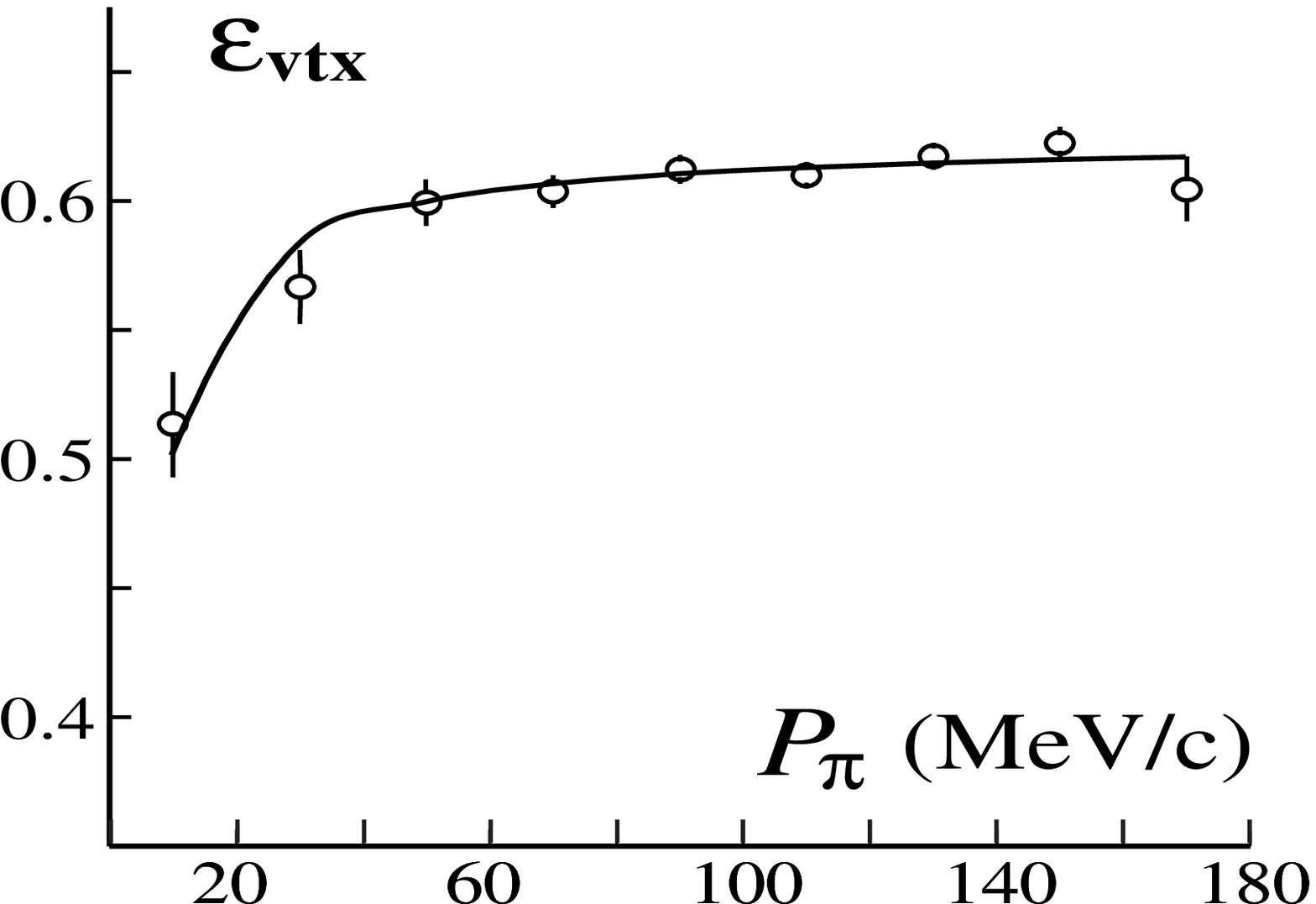}
\includegraphics{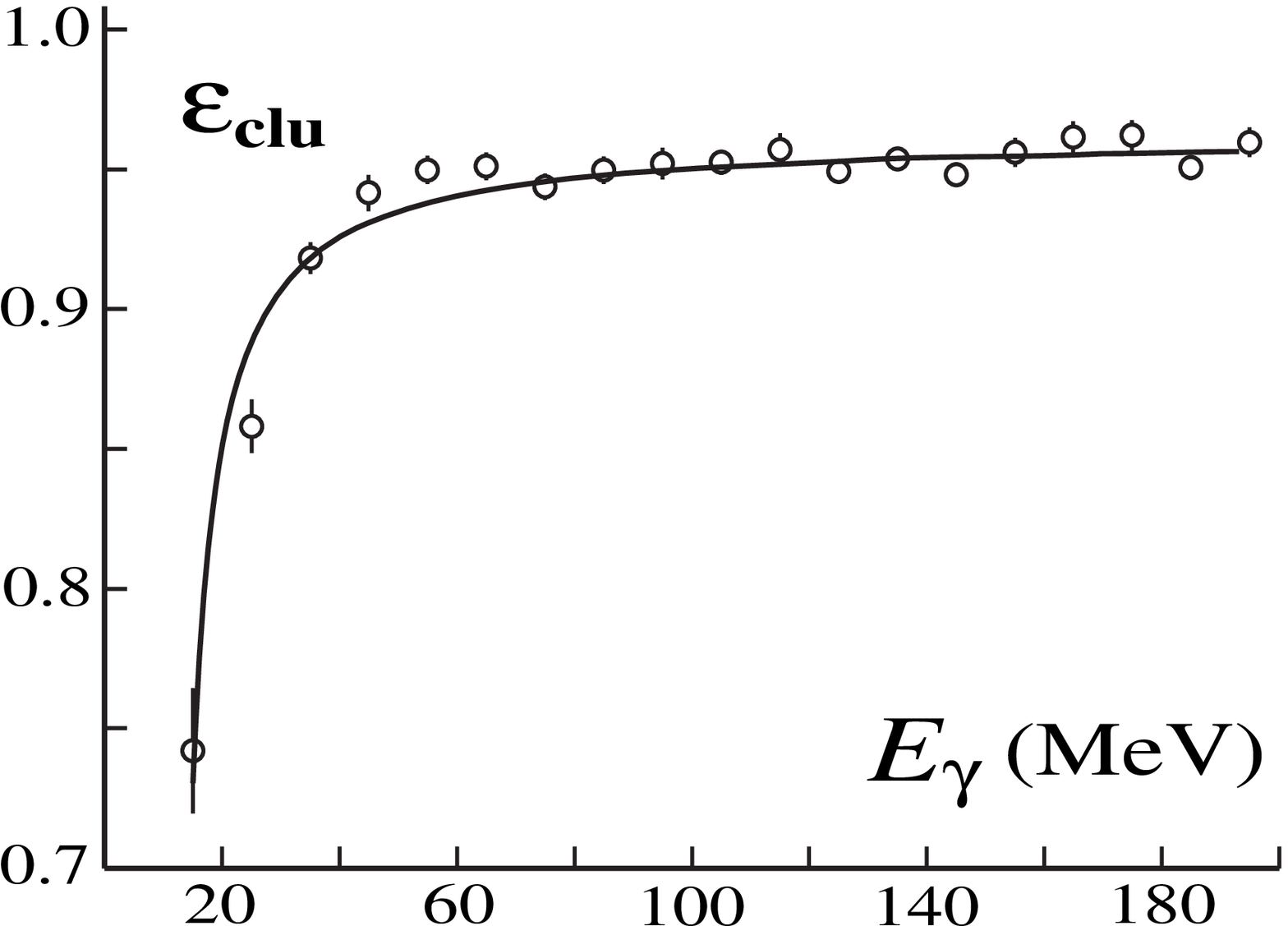}

\caption{Vertex efficiency as a function of the pion momentum in the lab frame 
(left). Single cluster efficiency corrected as described in the text as a function 
of the photon energy (right).\label{effic}}
\end{figure}

Where necessary, the efficiencies have been corrected for contamination of the 
control samples, as estimated by Monte Carlo simulation. For all samples the 
purities are $>95\%$. The efficiency $\epsilon_K$ has been found, within errors, 
to be the same for the two tags and kaon charges and its average value is equal to 
$0.466\pm 0.001_{stat}\pm 0.002_{syst}$. It is dominated by the geometrical 
acceptance (more than one third of the charged kaons decay before reaching the 
drift chamber). The effect of the above mentioned nuclear interactions of $K^\pm$ 
has been taken into account by properly weighting, to get $N_K$, the number of 
tagging kaons with a factor containing the probability for charged kaons to 
interact with the materials in front of the chamber. These weights have been 
extracted from data by using $K^\pm_{\pi2}$ and $K^\pm_{\mu2}$ samples: 
$w^+=0.9696\pm 0.0036$ and $w^-=0.9970\pm 0.0034$ respectively for positive and 
negative tags. \\
The efficiency for reconstructing the $K^\pm$-$\pi^\pm$ vertex, $\epsilon_{vtx}$, 
has been parameterized as a function of the charged pion momentum, as shown in 
Fig. \ref{effic} (left). After including the effect of the cut on the pion momentum,
the average value $\epsilon_v=0.539\pm 0.003_{stat} \pm 0.002_{syst}$ is obtained 
for the signal. The systematic error includes small differences between the two 
tags.\\
The cluster efficiency is evaluated from a control sample and corrected by using a 
Monte Carlo simulation to consider the effects of accidental clusters.
Fig. \ref{effic} (right) shows the single cluster efficiency $\epsilon_{clu}$ 
vs energy. The only appreciable source of the difference in $\epsilon_{\gamma}$ 
between the $K^\pm_{\mu2}$ and $K^\pm_{\pi2}$ tags results from the on-time cluster
requirement, due to the different cluster multiplicities in each event type. 
As average value on positive and negative kaon charge we find:
$\epsilon_{\gamma} = 0.645\pm 0.001_{stat}\pm 0.004_{syst}\ (0.625\pm 0.002_{stat}
\pm 0.005_{syst})$ in the case of $K^\pm_{\mu2}$ ($K^\pm_{\pi2}$) tag.\\
Finally, 
$\epsilon(\tau')=0.1621\pm 0.0010_{stat}\pm 0.0016_{syst}\ (0.1573\pm 0.0011_{stat}
\pm 0.0018_{syst})$ combining positive and negative kaons for $K^\pm_{\mu2}$ 
($K^\pm_{\pi2}$) tags. 

The residual dependence of our tagging procedure on the {\it tagged} kaon decay 
mode is expressed by the ratio $R_{tag}$ of the average (over all channels) 
tagging efficiencies (including trigger) and the values of the same efficiencies 
in the case of $\tau'$ signal. $R_{tag}$ is determined from Monte Carlo and the 
averages on the two kaon charges are: $R_{tag}= 1.058\pm 0.005\ (1.099\pm 0.009)$ 
for $K^\pm_{\mu2}$ ($K^\pm_{\pi2}$) tags, where 
the error includes systematic effects due to
small differences between data and Monte Carlo, as estimated on a 
set of variables to which $R_{tag}$ is found to be sensitive.

The $\tau'$ branching ratio is computed separately for $K^\pm_{\mu2}$ and
$K^\pm_{\pi2}$ tags and for the two signs of the kaon charge, from:
$${\rm BR}(K^\pm\rightarrow\pi^\pm\pi^0\pi^0) = {N(\tau')\over N_K}\times
\frac{1-f_{\rm bckgnd}(\tau')}{1-f_{\rm bckgnd}({\rm tag})}\times\frac{R_{tag}}
{\epsilon(\tau')}\times \frac{1}{[BR(\pi^0\rightarrow\gamma\gamma)]^2}$$
where BR$(\pi^0\rightarrow\gamma\gamma)$=0.98798$\pm$0.00032~\cite{pdg}. We find:
$$
{\rm BR}(K^+\rightarrow\pi^+\pi^0\pi^0) = (1.760\pm 0.024_{\rm stat}
\pm 0.017_{\rm syst})\times 10^{-2} \ (K^-_{\mu2} \ tag)
$$ 
\vspace{-8mm}
$$
{\rm BR}(K^-\rightarrow\pi^-\pi^0\pi^0) = (1.793\pm 0.024_{\rm stat}
\pm 0.026_{\rm syst})\times 10^{-2} \ (K^+_{\mu2}\ tag )
$$ 
$$
{\rm BR}(K^+\rightarrow\pi^+\pi^0\pi^0) = (1.769\pm 0.027_{\rm stat}
\pm 0.020_{\rm syst})\times 10^{-2} \ (K^-_{\pi2} \ tag)
$$ 
$$
{\rm BR}(K^-\rightarrow\pi^-\pi^0\pi^0) = (1.724\pm 0.027_{\rm stat}
\pm 0.027_{\rm syst})\times 10^{-2} \ (K^+_{\pi2} \ tag)
$$ 
yielding the average value:
$$
{\rm BR}(K^\pm\rightarrow\pi^\pm\pi^0\pi^0) = (1.763\pm 0.013_{\rm stat}
\pm 0.022_{\rm syst})\times 10^{-2} 
$$
where the final systematic error takes into account correlations. 
All contributions to the uncertainty are listed in Table \ref{summa}. 
The statistical error is dominated by the uncertainty in determining the 
efficiencies, due to the limited statistics of the data samples used in their 
evaluations. The stability of the result has been verified using data sets from 
different running periods. It should be pointed out that this measurement is 
fully inclusive of the radiative decays $K^\pm\rightarrow\pi^\pm\pi^0\pi^0\gamma$ 
(there is no cut on the radiated photon energy $E_\gamma$), since we require at 
least four (rather than exactly four) on-time clusters in the signal selection. 
An upper limit to the contribution of this decay is estimated to be 0.1$\%$ of the 
selected events, by taking into account possible differences in the efficiency 
for the two channels.
\begin{table}
\begin{center}
\begin{tabular}{|l|c|}
\hline
\hline
Source of uncertainty & Fractional error ($10^{-3}$)\\ \hline
$N(\tau')/N_K$ statistics & $3.5$ \\ \hline
Charged kaons nuclear interaction probability & $3.5$ \\ \hline 
Charged kaon reconstruction/identification efficiency & $5.2$ \\ \hline
Vertex reconstruction efficiency & $6.6$ \\ \hline
Cluster algorithm efficiency & $2.3$ \\ \hline
Four-cluster acceptance & $3.6$ \\ \hline
On-time requirement for clusters & $7.5$ \\ \hline
Total energy cut & $1.4$ \\ \hline
Background subtraction & $1.1$ \\ \hline
Ratio of tag efficiencies $R_{tag}$ & $6.6$ \\ \hline 
$BR(\pi^0\rightarrow\gamma\gamma)^2$ & $0.7$ \\ \hline 
\end{tabular}
\end{center}\vglue3mm
\caption{Summary of all contributions to the total error on
$BR(K^\pm \rightarrow \pi^\pm \pi^0 \pi^0)$. \label{summa}}
\end{table}

\ack
We thank the DA$\Phi$NE team for their efforts in maintaining low background running 
conditions and their collaboration during all data-taking. 
We want to thank our technical staff: 
G.F.Fortugno for his dedicated work to ensure an efficient operation of 
the KLOE Computing Center; 
M.Anelli for his continous support to the gas system and the safety of the
detector; 
A.Balla, M.Gatta, G.Corradi and G.Papalino for the maintenance of the
electronics;
M.Santoni, G.Paoluzzi and R.Rosellini for the general support to the
detector; 
C.Pinto (Bari), C.Pinto (Lecce), C.Piscitelli and A.Rossi for
their help during major maintenance periods.
This work was supported in part by DOE grant DE-FG-02-97ER41027; 
by EURODAPHNE, contract FMRX-CT98-0169; 
by the German Federal Ministry of Education and Research (BMBF) contract 06-KA-957; 
by Graduiertenkolleg `H.E. Phys. and Part. Astrophys.' of Deutsche Forschungsgemeinschaft,
Contract No. GK 742; 
by INTAS, contracts 96-624, 99-37; 
and by TARI, contract HPRI-CT-1999-00088.

\end{document}